\documentclass[a4paper, 12pt]{article}

\usepackage{amsmath}
\usepackage{amsfonts}
\usepackage{amssymb}
\usepackage{graphicx}
\title{{\textsc{Are We Living in a Higher Dimensional Universe?}}}
\author{\textbf{$\hbox{V H Satheesh Kumar}^{\star}$} and \textbf{{$\hbox{P K Suresh}^{\dagger}$}} \\ School of Physics, University of Hyderabad \\ Hyderabad 500 046, India. \\ $^\star$ \textit{vhsatheeshkumar@yahoo.co.uk}, $^\dagger$ \textit{pkssp@uohyd.ernet.in}}
\begin{document}
\date{}
\maketitle

\begin{abstract}
It is a brief review \footnote{Following the UN's invitation to celebrate the World Year of Physics in 2005, this is our humble tribute to Albert Einstein and his work which has drastically changed the way we look at Universe.} of the physical theories embodying the idea of extra dimensions, starting from the pre-historic times to the present day. Here we have classified the developments into three \textit{eras}, such as Pre-Einstein, Einstein and Kaluza-Klein. Here the  views and flow of thoughts are emphasized rather rigorous mathematical details. Majour developments in Quantum field theory and Particle physics are outlined. Some well known higher dimensional approaches to unification are discussed. This is concluded with some examples for visualizing extra dimensions and a short discussion on the cosmological implications and possible existence of the same.
\end{abstract}
\newpage
\section{Introduction}

Everyday experience and intuition tells that the space in which we live and move is three dimensional. Because we can go front and back, left and right or up and down, that is it! We have observed that our physical laws require no more than the three spatial dimensions to describe how a car works or why stone falls back when thrown up. Then, why to bother about extra-dimensions?\footnote{A beautiful account of extra dimensions can be found in Sanjeev S. Seahra's PhD Thesis at University of Waterloo. Here we have adopted a similar line of argument.}

The study of extra-dimensions was started at first place to address some of the issues like unification of the fundamental forces in nature, quantization of gravitational interaction, Higgs mass hierarchy problem and cosmological constant problem \textit{etc}. Perhaps by studying  higher-dimensional theories, we can explain some of the fundamental questions like, why are there only six types of quarks and leptons? or why are neutrinos massless(massive)? May be such a theory can give us new rules for calculating the masses of fundamental particles and the ways in which they affect one another. Or perhaps there may be some other arena in the physical world where extra dimensional space is required for explanation? Our life would not have been that interesting if we were not worrying about extra-dimensions!

\section{A Brief History}

\subsection{The Pre-Einsteinian Era}
 
Our perception of space has always been connected with our knowledge of geometry. Till the end of the nineteenth century, everybody believed that the space has got only three dimensions. Phythagorus did not even think of extra dimensions and Aristotle said three was all that space could have, for Euclid it was very obvious and Ptolemy even proved that the fourth is impossible! In the seventeenth Century, Ren\'e Descartes (1595-1650 AD) developed a method to graph points and lines on the plane by labeling points by a pair of numbers representing distances from some specified point along two orthogonal directions. This coordinate system is now commonly referred to as Cartesian in honor of Descartes. 

By the way, what is dimension? To be precise, ``the dimension of an object is a topological measure of the size of its covering properties." For easier understanding, the dimension of a space can thus be thought of as the number of coordinates needed to specify the location of a point in that space.  More generally, in $d$-dimensional Euclidean geometry, points labeled by a set of $d$ numbers representing distance from a specified point in $d$ mutually orthogonal directions. This served as  the physicist's idea of space. 

In the earlier developments of classical physics there was no ambiguity in the concept of space. It was just three dimensional Eucledian space whose properties seemed to be independent of anything and everything. When he wrote down his three axioms of motion, Newton defined time implicitly with the statement, \textit{Absolute, true, and mathematical time, of itself, and from its own nature, flows equably without relation to anything external} \cite{1}. He set forth this definition so that he could apply his mathematical formalism and not worry that the quantity in his equations might be affected by motion. Actually Newton was able to synthesize the earlier Galilean laws which govern the terrestrial motion and Kepler laws which govern the planetary motion with his law of universal gravitation. This law along with his laws of motion can explain all kinds of motion, terrestrial or celestial. This was a remarkable mile-stone in the history of physics. A similar such great synthesis was achieved by James Clark maxwell in 1873. He formulated all the laws of electricity and magnetism in a single set of four equations popularly know as Maxwell's equations. 

How do we know that our universe consists of only three spatial dimensions? Consider the following geometrical argument\cite{2}. There are five and only five regular polyhedra: the cube, dodecahedron, icosahedron, octahedron, and tetrahedron, as was proved by Euclid  in the last proposition of \textit{The Elements}. In 1750, the mathematician Leonhard Euler demonstrated an important relation between the number of faces $F$, edges $E$, and  vertices $V$ for every regular polyhedron: $V - E + F = 2$. We can see for ourself that all the above mentioned regular polyhedra satisfy this relation. Only these five solids satisfy this relationship - no more, no less. This is only in the case of three dimensions. But later curious mathematicians have generalized Euler's relationship to higher dimensional spaces ($V-E+F-C=0$; where $C$ is number of cells) and have come up with some interesting results. In a world with four spatial dimensions, for example, we can construct only six regular solids, \textit{viz} pentatope, tesseract, hexadecachoron, icositetrachoron, hecatonicosachoron and hexacosichoron. In higher dimensions (5, 6, 7 ....) there are only three regular polytopes in any particular dimensions! These three regular polytopes are the equivalent of the tetrahedron, cube, and octahedron in 3 dimensions, they are normally called the $n$-simplex (hypertetrahedron), $n$-cube (hypercube), and $n$-crosspolytope (hyperoctahedron) respectively where $n$ stands for the number of dimensions. Now, after all this, if our familiar world were not three-dimensional, geometers would not have found only five regular polyhedra after 2,500 years of searching. They would have found six in the case of four spatial dimension, or perhaps only three if we lived in a 5-D universe. Instead, we know of only five regular solids. And this suggests that we live in a universe with, at most, three spatial dimensions. Such a beautiful argument! 

\begin{figure}
   \centering
   \includegraphics[scale=0.6]{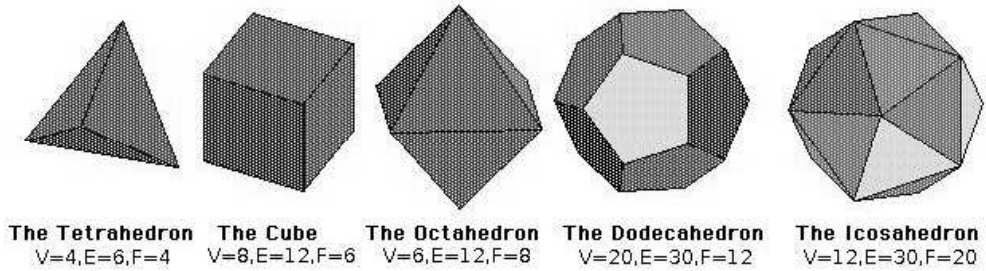}
   \caption{The five and only five regular polyhedra, also know as Platonic solids.}
   \label{fig:PlatonicSolids}
\end{figure}

In about 300 BC Euclid wrote \textit{The Elements}, in which he stated five postulates on which all his theorems were built. It is clear that the fifth postulate (parallel axiom) is different from the other four. It did not satisfy Euclid and he tried to avoid its use as long as possible - in fact the first 28 propositions of The Elements are proved without using it. Since then many mathematicians have been trying to prove Euclid's parallel axiom from the first four but in vain. To name a few, Proclus (410-485), Playfair, Wallis, Girolamo Saccheri, Lambert and Legendre. By this time elementary geometry was engulfed in the problems of the parallel postulate. D'Alembert, in 1767, called it \textit{the scandal of elementary geometry}. The first person to really come to understand the problem of the parallels was Gauss. He began work on the fifth postulate in 1792 while only 15 years old, at first attempting to prove the parallels postulate from the other four. Later Bolyai and Lobachevsky independently came up with new geometries. But it was actually Bernhard Riemann who gave the touch up to non-Eucledian geometries and developed a formalism to describe geometry of arbitrary dimensional manifold. Riemann, who wrote his doctoral dissertation under Gauss's supervision, gave an inaugural lecture on 10 June 1854 in which he reformulated the whole concept of geometry which he saw as a space with enough extra structure to be able to measure things like length. This was to have a profound influence on the development of a wealth of different geometries. That is how the concept of curved and n-dimensional spaces crept into the realm of mathematics. This work opened up new vistas in Mathematics and gave a death blow to Euclidean prejudice which had prevailed for over 2000 years! Unfortunately they did not connect their mathematical fantasies to any physical reality. Following Riemann's work, some mathematicians of the era who speculated about the idea of higher dimensional universe were Clifford and Hermann von Helmholtz. But the undisputed king of popularizing extra dimensions in the late 1800s was a mathematician Charles Howard Hinton. All these stuff served as fodder for various artistic and intellectual fancy but not in a quantitative way.

\subsection{The Einsteinian Era}

The nature of \textit{space} and \textit{time} has been an essential component of every theory since the beginning of physics. All was fine until the Maxwell's equations predicted the velocity of electromagnetic wave. This was same in all directions and completely voilated the Galilean law of addition of velocities. The cracks started appearing in the Newtonian edifice when Maxwellian physics could not be accommodated in the Newtonian framework. In 1905 the special relativity  emerged naturally in an attempt to save Electrodynamics from Classical mechanics. According Einstein\cite{3}, the speed of light is constant in all inertial frames. This blurred the distinction between \textit{space} and \textit{time} and one could be expressed in terms of the other.  Measurements of \textit{time} and \textit{space} are intimately connected with motion; time intervals are dilated and space intervals contracted by a factor which depends upon the relative velocity between two given frames of reference. Moreover, if one carries Einstein's reasoning beyond the simple ideas of time dilation and length contraction, one is forced to demote \textit{time} from its status as an independent entity and thereby unify it with space.  Time, then, becomes simply one dimension of a four-dimensional spacetime. Hermann Minkowski summed up all this and gave the geometric picture of spacetime in 1909. This gave the mathematical beauty to Einstein's Spacial Relativity. This geometric interpretation of $t$ as the fourth dimension  was very difficult for all to digest as, till then, time and space were treated on different footings. In this picture of spacetime, all the Maxwellian equations took a very simple but elegant form. This resulted in happy ending of Newton-Maxwell conflict. Actually this sent an unheard message that  ``some things that are complicated in lower dimensions are sometimes simpler in higher ones!!!"

Einstein later formulated the theory of general relativity, which extends relativity to include non-uniform motion and gravitation. To accomplish this feat, Einstein had to borrow the idea of curved space from Riemann. In this scenario the curvature is determined by the strength of the gravitational field. One consequence of this reasoning is gravitational time dilation, that is, the phenomenon that time flows more slowly in the presence of a gravitational field than in its absence. Thus the flow of time is no longer independent of something external. Today, general relativity is the accepted theory of gravity, and the concept of spacetime has become rooted in the minds of physicists. By this, the 4-dimensional (three space and one time) nature of the Universe was put on firm theoretical base.

\subsection{The Kaluza-Klein Era}

In 1914 Gunnar Nordstrom discovered that he could unite the physics of 	electromagnetism and gravity by postulating the existence of a fourth spatial dimension. However, recall that general relativity did not exist yet, so Nordstrom had not produced the correct theory of gravity, but only an approximation.

Actually the real story starts from here! By this time, there were two well tested field theories, Maxwell's Electromagnetic theory and Einstein's General Relativity. But the problem was to fuse them. There were already openions that the unification of Maxwell's and Einstein's formalisms are possible only in a higher dimensional theory. The man widely credited for this approach to unification was Theodor Kaluza.

In his 1921 paper, Kaluza\cite{4} considered the extension of general relativity to five dimensions. He assumed that the 5-dimensional field equations were simply the higher-dimensional version of the vacuum Einstein equation, and that all the metric components were independent of the fifth coordinate. The later assumption came to be known as the \textit{cylindrical condition}. This resulted in something remarkable: the fifteen higher-dimension field equations naturally broke into a set of ten formulae governing a  tensor field representing gravity, four describing a vector field representing electromagnetism, and one wave equation for a scalar field.  Furthermore, if the scalar field was constant,  the vector field equations were just Maxwell's equations in vacuo, and the tensor field equations were the 4-dimensional Einstein field equations sourced by an electromagnetic field. In one fell swoop, Kaluza had written down a single covariant field theory in five dimensions that yielded the four dimensional theories of general relativity and electromagnetism.

But there were problems with Kaluza's theory, not the least of which was the nature of the fifth dimension.  Before Minkowski, people were aware of time, they just had not thought of it as a dimension.  But now, there did not seem to be anything convenient that Kaluza's fifth dimension could be associated with. Moreover, there was no explanation given for Kaluza's \textit{ad hoc} assumption that none of the fields in the universe should vary over the extra dimension (\textit{the cylindrical condition}). 

In 1926 Oskar Klein \cite{5} provided an explanation for Kaluza's assumption, namely that if the extra dimension forms a circle much smaller than the distance scales we commonly observe. The theory of gravity on a compact space-time is called \textit{Kaluza-Klein theory}. Kaluza-Klein theory still had certain problems in its interpretation as a theory which united gravity and electromagnetism, most notably that it predicted a unit of charge which was many times smaller than the electrons charge. As a result, many physicists left the idea of extra dimensions for the realm of curiosity, until 1970s. 

Kaluza's and Klein's 5-dimensional version general relativity that contains both electromagnetism and 4-dimensional general relativity, although flawed, is an example of such an attempt to unite the forces of nature under one theory. It led to glaring contradictions with experimental data, but some physicists felt that it was on the right track, that it in fact didn't incorporate enough extra dimensions! This led to modified versions of Kaluza-Klein theories incorporating numerous and extremely small extra dimensions.

Among the modern Kaluza-Klein theories \cite{6}, the three main approaches to higher dimensional unification are
\begin{enumerate}
\item{\textit{Compactified} approach, where the extra-dimensions are compactified and unobservable on experimentally accessible energy scales.}
\item{\textit{Projective} approach, where extra coordinates are not physically real.}
\item{\textit{Non-compactified} approach, in which extra dimensions are not necessarily length-like or compact.}
\end{enumerate}

All right, now we try to give a similar argument as we did in first part of this section and try if we can prove the presence of extra dimension(s). Let us suppose our universe actually consists of four spatial dimensions. What happens? Since relativity tells us that we must also consider time as a dimension, we now have a space-time consisting of five dimensions. 

To the best available measurements, gravity follows an inverse square law; that is, the gravitational attraction between two objects rapidly diminishes with increasing distance. For example, if we double the distance between two objects, the force of gravity between them becomes 1/4 as strong; if we triple the distance, the force becomes 1/9 as strong, and so on. A five-dimensional theory of gravity introduces additional mathematical terms to specify how gravity behaves. These terms can have a variety of values, including zero. If they were zero, however, this would be the same as saying that gravity requires only three space dimensions and one time dimension. The fact that the planetary positions (except Mercury, which requires corrections due to general relativistic effects) and different epochs of solar eclipse agree within a few seconds of their predicted times is a wonderful demonstration that we do not need extra-spatial dimensions to describe motions in the Sun's gravitational field. From the above  physical arguments, we can conclude, not surprisingly, that space is three-dimensional on scales ranging from that of everyday objects to at least that of the solar system, i.e. 1 to $10^{17} cm$. If this were not the case, gravity would function very differently than it does, for example, eclipses would not have happened in time! Yet another beautiful argument, but what about the behavior of gravity at scales less than $1 cm$ or greater than $10^{17} cm$. Only the future experiments \cite{7} can answer this.

\section{The Resurrection}

The Kaluza-Klein theory was the first promising theory to combine gravity and electromagnetism. Unfortunately this approach never really caught on and was soon buried by the onrush of theoretical work on quantum theory. It was not until work on supergravity theory began in 1975 that Kaluza and Klein's method drew renewed interest. Physicists started searching for Kaluza and Klein's papers in their dark store rooms! Its time had finally come. Now we see what were the new ideas that sprouted in physicist's minds during this pause.

Starting around 1927, Paul Dirac unified quantum mechanics with special relativity and attempts were made to apply quantum mechanics to fields rather than single particles, resulting in what are known as quantum field theories. Early workers in this area included Dirac, Pauli, Weisskopf, and Jordan. This area of research culminated in the formulation of quantum electrodynamics by Feynman, Dyson, Schwinger, and Tomonaga during the 1940s. Quantum electrodynamics is a quantum theory of electrons, positrons, and the electromagnetic field, and served as a role model for subsequent quantum field theories.

Building on pioneering work by Schwinger, Higgs, Goldstone and others, Sheldon Glashow, Abdus Salam and Steven Weinberg independently showed in 1968, how the weak nuclear force and quantum electrodynamics could be merged into a single electroweak force. The Standard Model of particle physics describes the electromagnetic force and the weak nuclear force as two different aspects of a single electroweak force. 

The theory of quantum chromodynamics as we know it today was formulated by Politzer, Gross and Wilzcek in 1975. This explains the strong interaction and forms an important part of the standard model of particle physics. 

A theory that unifies the strong interaction and electroweak interaction is called
Grand Unified Theory, or GUT. These models generically predict the existence of topological defects such as monopoles, cosmic strings, domain walls, and others. None have been observed and their absence is known as the monopole problem in cosmology. GUT models also generically predict proton decay, although current experiments still haven't detected proton decay.

The Standard Model of particle physics is a theory which describes the strong, weak, and electromagnetic interactions, as well as the fundamental particles that make up all matter. To date, almost all experimental tests of the three forces described by the Standard Model have agreed with its predictions. However, the Standard Model is not a complete theory of fundamental interactions, primarily because it does not describe gravity. Now it is time to unify gravity with the other fundamental interactions of the nature.

\section{Unification}

The two pillars of modern physics, general relativity and quantum mechanics, are the most well-tested and experimentally verified theories in existence. Each have been rigorously examined for the better part of the last century, and have stood up amazingly well under all scrutiny. A problem arises, however, when one attempts to probe the depths of ultramicroscopic scales that quantum mechanics describes so well. General relativity gives us a picture of smooth spacetime, but according to quantum mechanics, this is not the case at very small levels. The main thrust of quantum mechanics, the uncertainty principle, says that on such small scales, properties of objects that we take for granted, such as velocity, position, momentum, etc, are all uncertain and probabilistic. In this realm, general relativity absolutely must be replaced by a quantum theory of gravity. To devise such a theory, it will probably be necessary to incorporate the discontinuity of spacetime, or to do away with spacetime altogether! One can achieve this by going for a higher dimensional theory in the same light as the Minkowski 4-dimensions unified electromagnetism. Now we discuss some important higher-dimensional approaches to get the Theory of Everything, such as,
\begin{enumerate}
\item Supersymmetry and Supergravity
\item String Theory
\item Space-Time-Matter Theory
\item Braneworlds
\end{enumerate}

\subsection{Supersymmetry and Supergravity}
As the years rolled on physicists discovered that it no longer seemed that the 5-dimensional Kaluza-Klein model was a viable candidate for a theory of everything. Because the strong and weak interactions require more degrees of freedom than a 5-metric could offer. However, the way in which to address the additional requirements of modern physics is not hard to imagine, one merely has to further increase the dimensionality of theory until all of the desired gauge bosons are accounted for.  Then how many dimensions do we need to unify modern particle physics with gravity via the Kaluza-Klein mechanism? The answer is at least eleven, which was shown by Witten \cite{8} in 1981. This result prompted the consideration of an 11-dimensional extension of general relativity.   But such a construction cannot really be the theory of everything; there are no fermions in the model, only gauge bosons. This is where the idea of supersymmetry enters in the picture, which is a hypothesis from particle physics that postulates that the Lagrangian of the universe  that is, the action principle governing absolutely everything  is invariant under the change of identities of bosons and fermions. This means that every boson and fermion has a superpartner with the opposite statistics; i.e., the graviton is coupled with the gravitino, the photon with the photino, etc. The gravitino has the remarkable property of mathematically moderating the strength, of various kinds of interactions involving the exchange of gravitons. The photino, cousin of the photon, may help account for the ``missing mass" in the universe.  

When supersymmetry is coupled with general relativity, one has supergravity which is a theory of 11 dimensions. There are several reasons why that eleven dimensions are attractive. In 1978, Nahm showed that supergravity was consistent with asymmetry operations on an n-sphere; i.e., rotations. Also in 1978, Cremmer, Julia, and Scherk demonstrated that the supergravity Lagrangian in eleven dimensions was in some sense unique; one did not have any wiggle room to alter the higher-dimensional action principle in a fundamental way. Finally, in 1980, Freund and Rubin showed that the theory could be dynamically compactified in a manner consistent with the higherdimensional field equations. However, all was not well with the model. For chiral fermions in 4 dimensions were not forthcoming, and the theory suffered from the renormalization problems. The former was alleviated by going down to ten dimensions at the price of losing the uniqueness of the 11-dimensional theory, but the latter was a harder nut to crack. Supergravity theories are believed to be the only consistent theories of interacting spin 3/2 fields. These problems caused segments of the community to move on from supergravity to another higher-dimensional programme called String Theory.
 
\subsection{String Theory}

In the 1960s, particle physicists reached towards something called a \textit{dual resonance} model in an attempt to describe the strong nuclear force. The dual model was never that successful at describing particles, but it was understood by 1970 that the dual models were actually quantum theories of relativistic vibrating strings and displayed very intriguing mathematical behavior. Dual models came to be called \textit{string theory}\cite{9} as a result. 

In the early 1970's when people tried to find a renarmalizable theory which incorporates all the fundamental interactions in the same spirit of GUT, their integrals shot up to infinity. One big problem was that classical gravitational waves carry spin 2, so one should assume that a graviton has spin 2.  But the string theory that was once proposed for the strong interactions contained a massless particle with spin 2. This was sheer coincidence and by 1974 people started speculating whether string theory can be a theory of quantum gravity?

One potentially successful theory to bridge the gap between quantum mechanics and general relativity is string theory. The basics of string theory assert that the fundamental components of the matter are not particles, but tiny, one-dimensional vibrating strings. These strings are as small as the Plank length, $10^{-35}$ m, and as they vibrate, their resonant frequencies determine the properties, such as mass or charge of the particles they constitute. The key behind replacing point particles by strings with some dimension is that this can remove the mathematical absurdities arising in the equations of general relativity when applied to such small scales. What results is a theory that embodies all aspects of quantum mechanics and general relativity without any incompatibilities!

The general solution to the relativistic string equations of motion looks very similar to the classical nonrelativistic case. But unlike a guitar string  it isn't tied down at either end and so travels freely through spacetime as it oscillates. The string above is an open string, with ends that are floppy.  For a closed string, the boundary conditions are periodic, and the resulting oscillating solution looks like two open string oscillations moving in the opposite direction around the string. These two types of closed string modes are called right-movers and left-movers.

When we add quantum mechanics by making the string momentum and position obey quantum commutation relations, the quantized string oscillator modes wind up giving representations of the Poincar\'e group, through which quantum states of mass and spin are classified in a relativistic quantum field theory. So this is where the elementary particles arise in string theory.

In the generic quantum string theory (includes only bosons), there are quantum states with negative norm, also known as \textit{ghosts}. This happens because of the minus sign in the spacetime metric. So there ends up being extra unphysical states in the string spectrum. In 26 spacetime dimensions, these extra unphysical states wind up disappearing from the spectrum. Therefore, bosonic string quantum mechanics is only consistent if the dimension of spacetime is 26. The reasons for higher dimensions in string theory are actually quite different than in Kaluza-Klein supergravity; they are required for the internal consistency of the model, as opposed to direct unification of the fundamental forces.
 
The bosonic string propagating in flat 26-dimensional spacetime  can give rise to four different quantum mechanically consistent string theories, depending on the choice of boundary conditions used to solve the equations of motion. The choices are divided into two categories:
\begin{enumerate}
\item  Are the strings open (with free ends) or closed (with ends joined together in a loop)?
\item   Are the strings orientable (you can tell which direction you're traveling along the string) or unorientable (you can't tell which direction you're traveling along the string)?
\end{enumerate} 
There are four different combinations of options, giving rise to the four bosonic string theories
     
The bosonic string theories are all unstable because the lowest excitation mode, or the ground state, is a \textit{tachyon}, a hypotheticle particle which moves at speed greater than that of light. It's just as well that bosonic string theory is unstable, because it's not a realistic theory to begin with. The real world has stable matter made from fermions that satisfy the Pauli Exclusion Principle where two identical particles cannot be in the same quantum state at the same time.   Adding fermions to string theory introduces a new set of negative norm states or ghosts to add to the earlier misery! String theorists learned that all of these bad ghost states decouple from the spectrum when two conditions are satisfied: the number of spacetime dimensions is 10, and the theory is supersymmetric, so that there are equal numbers of bosons and fermions in the spectrum. The resulting consistent string theories, popularly know as \textit{Superstring theories} \cite{21}, can be described in terms of the massless particle spectrum and the resulting number of spacetime supersymmetry charges. None of these theories suffer from the tachyon problem that plagues bosonic string theories and  all of them  contain graviton along with its supersymmetric partner \textit{gravitino} which has spin 3/2. There are five different kinds of superstring theories.

Superstring theories are not just theories of one-dimensional objects. There are higher dimensional objects in string theory with dimensions from zero to nine, called \textit{p-branes}. In terms of branes, what we usually call a membrane would be a two-brane, a string is called a one-brane and a point is called a zero-brane. A special class of p-branes in string theory are called \textit{d-branes}. Roughly speaking, a d-brane is a p-brane where the ends of open strings are localized on the brane.

At one time, string theorists believed there were five distinct superstring theories. But now it is known that this naive picture was wrong, and that the the five superstring theories are connected to one another as if they are each a special case of some more fundamental theory. In the mid-nineties it was learned that various string theories floating around were actually  related by duality transformations known as \textit{T-duality} and \textit{S-duality}. These ideas have collectively become known as \textit{M-theory}, where M is for membrane, matrix, or mystery, depending on your point of view!

\subsection{Space-Time-Matter Theory}

In Kaluza-Klein theories our 4 dimensional universe is believed to be embedded in a higher-dimensional manifold. Here, in this section, we consider Wesson's \textit{Space-Time-Matter} (STM) theory\cite{10}. This theory is distinguished from the classical Kaluza-Klein theory by the fact that it has an noncompact fifth dimension and that it is empty viewed from 5D and sourceful viewed from 4D. Because of this, the STM theory is also called \textit{induced matter theory} and the effective 4D matter is also called \textit{induced matter}. That is, in STM theory, the 5D manifold is Ricci-flat while the 4D hypersurface is curved by the 4D induced matter.

In the early 1990s, STM theory made its first appearance .To make STM consistent with the non-observation of the fifth dimension, one needs to make one of several possible assumptions about the nature of the extra dimension.  One idea stems from the state of affairs before Minkowski.  Back then, time was not viewed as a dimension for two reasons: first, the existing three dimensions were spacelike, and it took a significant leap of imagination to put time on the same footing;  second, the size of the dimension-transposing parameter $c$, the speed of light, made practical experimental verification of the dimensional nature of time difficult.  So, one way to reconcile a large fifth dimension is to suppose that it is neither timelike or spacelike, and that its scale prevents its signature from showing up in contemporary experiments.  But if the fifth dimension is not temporal or spatial, what is it? An interesting suggestion is that it is mass-like; i.e., one's position or momenta in the fifth dimension is related to one's mass. This idea brings a certain symmetry to ideas about dimensional quantities in mechanics, especially when we recall that physical units are based on fundamental measures of time, space, and mass. We commonly associate geometric dimensions with the first two, but not the third. The appropriate dimension transposing parameter between mass and length is $G/c^2 \sim 7.4 \times 10^{-28}$ metres per kilogram. Unlike $c \sim 3.0 \times 10^8$ metres per second, this number is small, not large. The implication is that a particle would have to exhibit large variations in mass to generate significant displacements in the extra dimension. Since such variations in mass are experimentally known to be small, the world appears to be 4-dimensional.

\subsection{Braneworlds}

An extremely popular alternative model involving extra dimensions is the so-called \textit{braneworld} scenario. This phenomenological model has been motivated by the work of Horava and Witten \cite{11}, who found a certain 11-dimensional string theory scenario where the fields of the standard model are confined to a 10-dimensional hypersurface, or brane. Actually the first proposal for using large (TeV) extra dimensions in the Standard Model with gauge fields in the bulk and matter localized on the orbifold fixed points came from Antoniadis\cite{12}, although the word brane was not used. In this picture the  non-gravitational degrees of freedom which are represented by strings whose endpoints reside on the brane and on the other hand, gravitational degrees of freedom in string theory are carried by closed strings, which cannot be tied-down to any lower-dimensional object. Hence, the main feature of this model is that the Standard Model particles are localized on a 3-brane, while gravity can propagate in 4+N dimensions called \textit{bulk}.

The central idea is that our visible, four-dimensional universe is entirely restricted to a brane in a higher-dimensional space, called the bulk. The additional dimensions may be compact. As one of its attractive features, the model can explain the \textit{weakness of gravity} relative to the other fundamental forces of nature. In the brane picture, the other three SM interactions are localized on the brane, but gravity has no such constraint and ``leaks" into the bulk. As a consequence, the force of gravity should appear significantly stronger on small say, sub-millimeter scales, where less gravitational force has ``leaked".

There are three distinct mechanisms by means of which the laws of (3+1) dimensional gravity can be obtained on a brane and they are discussed in brief below. Some what detailed discussion is given in reference \cite{13}
\\
\begin{itemize}
\item \textbf{Braneworlds with Compact Extra Dimensions} 

Here to obtain  (3+1) dimensional gravity on the brane the idea of KK compactification is combined with braneworld idea. This was proposed in 1998 by Arkani-Hamed, Dimopoulos and Dvali \cite{14} \cite{20} along with Antoniadis \cite{19} . This opens up new possibilities to solve the Higgs mass hierarchy problem and gives rise to new predictions that can be tested in accelerator, astrophysical and table-top experiments.  The fundamental scale of gravity and the UV scale of SM are around a few TeV.

\item \textbf{Braneworlds with Wrapped Extra Dimensions}

This phenomenon of localizing gravity was discovered by Randall and Sundrum \cite{15} in 1999. RS brane-worlds do not rely on compactification to localize gravity on the brane, but on the curvature of the bulk, sometimes called ``warped compactification". What prevents gravity from \textit{leaking} into the extra dimension at low energies is a negative bulk cosmological constant. There are two popular models. The first, called RS1, has a finite size for the extra dimension with two branes, one at each end. The second, RS2, is similar to the first, but one brane has been placed infinitely far away, so that there is only one brane left in the model. They also used their model to try to explain the hierarchy problem \cite{18} in particle physics. 

\item \textbf{Braneworlds with Infinite Volume Extra Dimensions}

This mechanism of obtaining (3+1) gravity on the brane is different from the earlier two as it allows the volume of the extra dimension to be infinite. This model was proposed in 2000 by G.R.Dvali, G.Gabadadze and M.Porrati \cite{16}. In this scenario the size of the extra dimensions does not need to be stabilized since the extra dimensions are neither compactified nor wrapped because of the presence of infinite-volume extra dimensions and hence gravity is modified at large distances. This gives rise to new solutions for late-time cosmology and acceleration of the universe which comes from type Ia supernovae observations. This can also explain dark energy problem and Cosmic Microwave Background.
\end{itemize}
\section{Visualizing Extra Dimensions}

So far we discussed many theories based on the idea of extra dimensions.In the context of modern unified field theory, these extra dimensions are, in a sense, internal to the particles themselves - a \textit{private secret} shared only by particles and the fields that act on them! These dimensions are not physically observable in the same sense as the three spatial dimensions we experience. Then how to imagine them?

\begin{figure}
\centering
\includegraphics[scale=1]{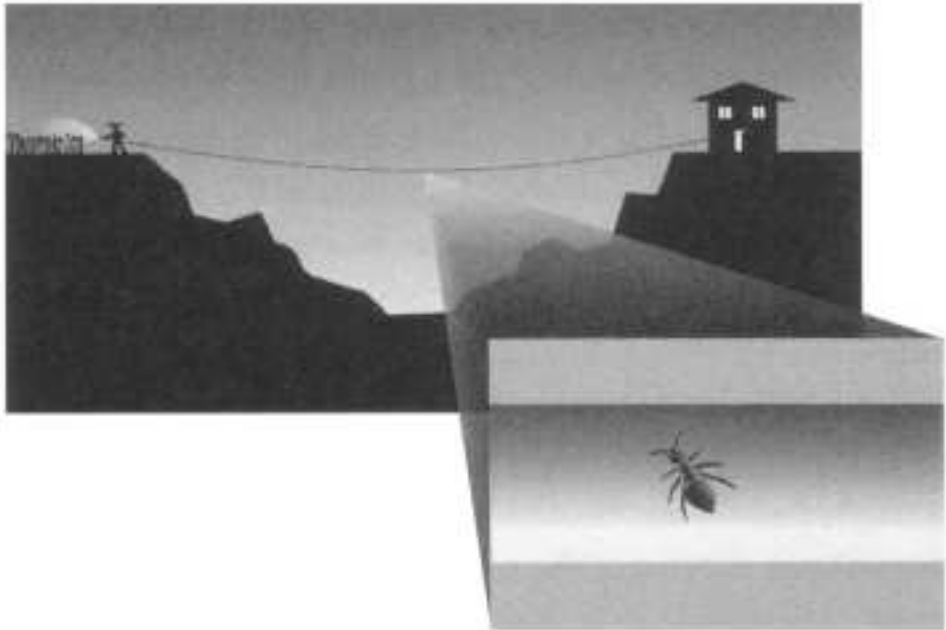}
\caption{From a distance the hose appears to be one-dimensional; close up, however, we see it as a two-dimensional surface. [Ref\cite{17}, p 186]}
\label{fig:ant-hose-small}
\end{figure}

In \textit{The Elegant Universe}\cite{17}, Brian Greene gives the following examples to visualize the extra dimensions. Imagine a hose of several hundred feet stretched out over a canyon, and imagine an ant crawling on the hose. From very far away, one sees the hose as a one-dimensional object: a line with no thickness. If viewing with binoculars, one may then notice that the ant, instead of being limited to travel in one dimension, has the ability to travel around the surface of the hose. Thus, magnifying the hose reveals another dimension. This previously unnoticed dimension is analogous to how a tiny and curled up dimension might exist in reality. It would be very small, much smaller than our current technology would enable us to measure, smaller than a billionth of a billionth of a meter, and its influence would obviously not be felt on any larger scales.

\begin{figure}
   \centering
   \includegraphics[scale=0.95]{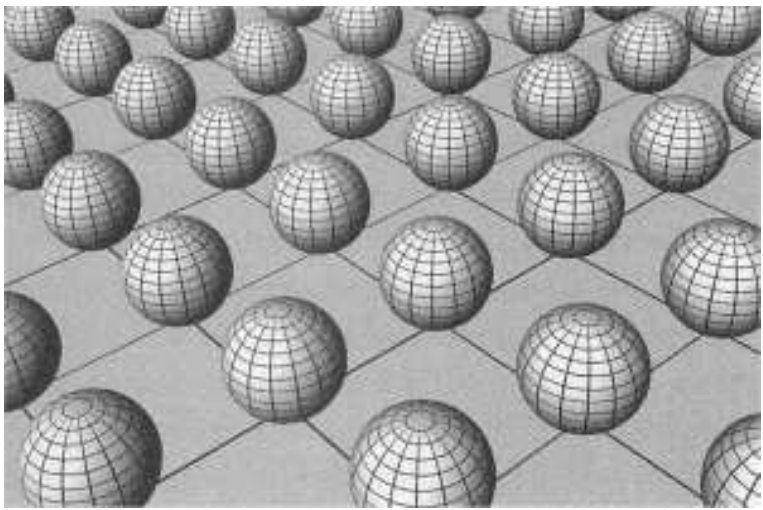}
   \caption{Two normal dimensions with 2 extra curled-up dimensions represented by spheres.[Ref\cite{17}, p 199]}
   \label{fig:spheregrid-small}
\end{figure}

To help visualize two tiny and curled up extra dimensions, we can return to the analogy of the ant on the hose, and how the hose has an extra dimension when we view it with magnification. Now, instead of a one-dimensional line, imagine a plane in two dimensions. Just as the hose has an extra dimension at each point, we can see an extra dimension in the form of a loop or a circle at each point on the plane. If we use spheres, we now have two curled up dimensions at each point on the plane. Remember that even though the image shows spheres only at intersections of grid lines, there would be spheres at every point on the plane. As you wave your hand through the air, it would pass through the normal three large dimensions, as well as these unimaginably tiny dimensions at each point along its path of motion, but only thing is you do not feel them!

\begin{figure}
   \centering
   \includegraphics[scale=0.75]{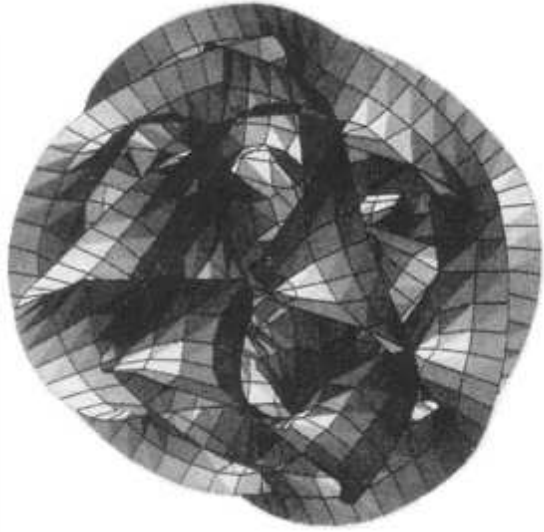}
   \caption{Two normal dimensions with 6 extra dimensions curled up in Calabi-Yau spaces [Ref\cite{17}, p 208].}
   \label{fig:calabi-yau-space3}
\end{figure}

\begin{figure}
   \centering
   \includegraphics[scale=0.73]{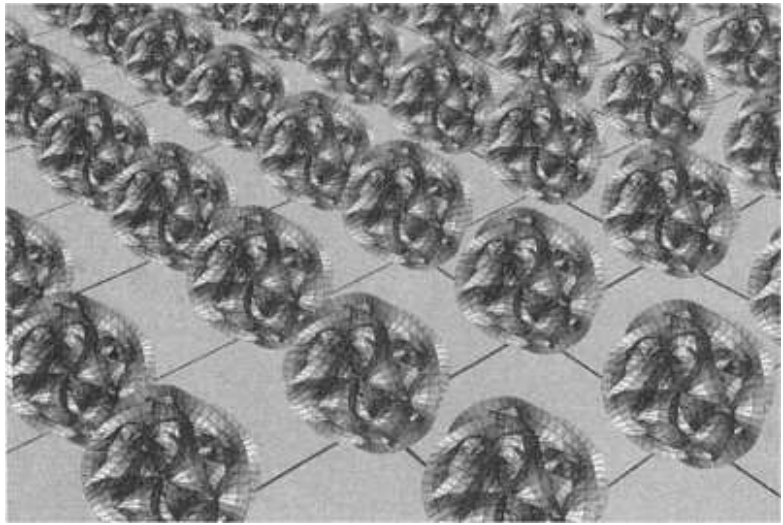}
   \caption{A representation of a 6-dimensional Calabi-Yau space [Ref\cite{17}, p 207].}
   \label{fig:calabi-yau-grid-small}
\end{figure}

What might a higher-dimensional form look like? The Calabi-Yau space[Fig.4] is a six-dimensional form that some of the above theories incorporate as the extra curled up dimensions. We can replace the sphere on our plane with the Calabi-Yau space [Fig.5] to visualize what these curled up dimensions may look like.

\section{Conclusion}

We have systematically revisited the concepts of dimensions and extra-dimensions over the centuries and the various theories built upon these ideas. Although we have not seen extra dimensions, some theories are mathematically consistent only if space has more number of dimensions. It therefore seems very plausible that there are extra dimensions of space, of course, we don't observe them in everyday life because they are small. We have a hard time visualizing more than three dimensions to space, but this is just a limitation of our imagination.

The assumtiom of extra dimensions which arose out of need in particle physics has its repercussions in cosmology too. This offers new explanations for cosmic microwave background, dark matter, dark energy, inflation, baryogenesis and cosmological constant.

However, the progress of physics lies in explaining more phenomena on the basis of theories that are constrained by standards of logic, conciseness and elegance. But none of these theories can be ruled out by the classical tests of relativity or results from astrophysics and cosmology. Then what is the fundamental test of physics which determines the existence of  these extra dimensions? 
\begin{itemize}
\item One thing is to see the behavior of gravity at low distances, because so far, this has been tested down to a distance of a centimeter only. In such a case, the radii of the extra dimensions need to be smaller than this. There is an exciting possibility that the upcoming submillimeter measurements of gravity will uncover some aspects of the above discussed theories.

\item The second way is to look for non-conservation of energy in some collider experiments. This missing energy can be related to the production of gravitons which escape into the extra dimensions. This  way to test the existence of extra dimensions may be possible with the next generation collider LHC at CERN. This will allow us to ultimately prove the theory of extra dimensions or not. 

\item The other way is to look for similar energy loss mechanisms in some astronomical sources. One such candidate is type II supernova, whose cooling is faster than the theoretical predictions. Such energy-loss mechanisms in supernovae can put bound on the number of extra dimensions. This can also be supplimented by the collider experiments.
\end{itemize} 

Discovery of extra dimensions would herald the first change in our view of spacetime since Einstein's theory of relativity. Perhaps it may solve some long standing problems in physics and astrophysics. But the same question haunts again ``Are we living in a higher dimensional universe?"
\pagebreak
\section*{Acknowledgement}
We are grateful to Ignatius Antoniadis and  Christos Kokorelis for some suggestions and Brian Greene for picture courtesy. We also acknowledge some valuable comments from Sergei Odinstov, Zhong Chao Wu, Lorenzo Iorio and Gerald Vones. We would like to thank Rizwan Ul Haq Ansari for useful discussions.


\begin{thebibliography}{999}
\bibitem{1} C.W.Misner, K.S.Thorne and J.A.Wheeler, \textit{Gravitation}, W.H.Freeman Publications, 1972.
\bibitem{2} S.Odenwald, Astronomy, November (1984).
\bibitem{3} A.Einstein \textit{The meaning of Relativity}, Princeton University Press, 1965.
\bibitem{4} Th.Kaluza, Sitzungsber. Preuss. Akad. Wiss. Berlin, Math. Phys. K1, 966 (1921).
\bibitem{5} O.Klein, Z. Phys. 37, 895 (1926).
\bibitem{6} J.M.Overduim and P.S.Wesson, Phys. Rept. 283, 303 (1997).
\bibitem{7} J.C.Long, H.W.Chan and J.C.Price, Nucl. Phys. B, 539, 23 (1999).
\bibitem{8} E.Witten, Nucl. Phys. B, 186, 412 (1981).
\bibitem{9} Official String Theory Web Site, http://www.superstringtheory.com/
\bibitem{21} M.B.Green, Scientific American, September (1986).
\bibitem{10} P.S.Wesson, \textit{Space - Time - Matter: Modern Kaluza-Klein Theory}, World Scientific Publishing Company, 1998.
\bibitem{11} P.Horava and E.Witten, Nucl. Phys. B 475, 94 (1996). hep-th/9603142.\\ P.Horava and E.Witten, Nucl. Phys. B 460, 506 (1996). hep-th/9510209.
\bibitem{12} I.Antoniadis, Phys. Lett .B 246, 317 (1990).
\bibitem{13} G. Gabadadze,  hep-ph/0308112.
\bibitem{14} N.Arkani-Hamed, S.Dimopoulos and G.R.Dvali, Phys. Lett. B 429, 263, (1998). hep-ph/9803315.
\bibitem{20} N.Arkani-Hamed, S.Dimopoulos, and G.R.Dvali, Phys. Rev. D 59,086004 (1999). hep-ph/9807344.
\bibitem{19} I.Antoniadis, N.Arkani-Hamed, S.Dimopoulos, and G.R.Dvali, Phys. Lett. B 436,257 (1998). hep-ph/9804398.
\bibitem{15} L.Randall and R.Sundrum, Phys. Rev. Lett. 83, 4690 (1999). hep-th/9906064. 
\bibitem{18}L.Randall and R.Sundrum, Phys. Rev. Lett. 83, 3370 (1999). hep-ph/9905221.
\bibitem{16} G.R. Dvali, G.Gabadadze and M. Porrati, Phys. Lett. B 485, 208 (2000). hep-th/0005016.
\bibitem{17} B.Greene, \textit{The Elegant Universe}, Vintage, 2000.
\end{thebibliography}
\end{document}